\documentclass[
groupedaddress,
reprint,
 amsmath,amssymb,
 aps, physrev,
pr4,
]{revtex4-2}

\usepackage{graphicx}
\usepackage{dcolumn}
\usepackage{bm}

\begin{document}

\preprint{APS/123-QED}

\title{Machine learning prediction of a chemical reaction over 8 decades of energy} 

\author{Daniel Julian}
\affiliation{Department of Physics and Astronomy, Stony Brook University, 11794 Stony Brook, NY, USA
}%

\author{Jes\'us P\'erez-R\'ios}
 \email{Contact author: jesus.perezrios@stonybrook.edu}
\affiliation{Department of Physics and Astronomy, Stony Brook University, 11794 Stony Brook, NY, USA
}%

\date{\today}

\begin{abstract}

Recent progress in machine learning has sparked increased interest in utilizing this technology to predict the outcomes of chemical reactions. The ultimate aim of such endeavors is to develop a universal model that can predict products for any chemical reaction given reactants and physical conditions. In pursuit of ever more universal chemical predictors, machine learning models for atom-diatom and diatom-diatom reactions have been developed, yet no such models exist for termolecular reactions. Accordingly, we introduce neural networks trained to predict opacity functions of atom recombination reactions. Our models predict the recombination of Sr$^+$ + Cs + Cs $\rightarrow$ SrCs$^+$ + Cs and  Sr$^+$ + Cs + Cs $\rightarrow$ Cs$_2$ + Sr$^+$ over multiple orders of magnitude of energy, yielding overall results with a relative error $\lesssim 10\%$. Even far beyond the range of energies seen during training, our models predict the atom recombination reaction rate accurately. As a result, the machine is capable of learning the physics behind the atom recombination reaction dynamics.

\end{abstract}

\maketitle


Machine learning (ML), within the context of artificial intelligence (AI), is transforming the current paradigm in the physical sciences, ushering in the big data era. In this emerging paradigm, the methodology involves using data directly to identify patterns and predict the behavior of complex systems. In this regard, the calculation of the outcome of a chemical reaction is one of the most demanding and needed tasks in modern chemical physics. As a result, considerable effort has been dedicated to using data-driven approaches in chemical physics to predict chemical reactions~\cite{Zhengkai2023,Tkatchenko,Mewly2021,chemistry2,chemistry4,chemistry5}. For instance, several methods have been developed to accelerate the calculation of multi-dimensional potential energy surfaces~\cite{Mewly2023,miksch2021strategies,rasheeda2022high,behler2007generalized,cui2016efficient,qu2018assessing,Xiangyue2023}, the essential element for reaction dynamics calculations. Similarly, the prediction of reaction barrier heights—the bottleneck to understanding many essential chemical reactions— is recently gaining more attention ~\cite{Andrade2023,barrier_1,barrier_2,barrier_3,barrier_4,barrier_5,barrier_6}. In the case of complex chemical reactions, machine learning is used to automate reaction networks in intricate chemical scenarios~\cite{smith2018less,vandermause2020fly,Reaction_network2,Reaction_network1,RN}. 

Despite significant progress, the main goal remains elusive: a universal chemical predictor. A universal predictor would take a set of reactants and physical properties (like temperature and density) and predict the products of the reaction and their branching ratios. It has been possible to predict the outcome of atom-diatom and diatom-diatom reactions~\cite{Koner2019,Dissociation,Panesi2023,Gu2023,Arnold2022,COO,Forrey2020,Forrey_2025} for a given set of reactants or across the isotopologue space~\cite{Dan2024}. The next level of complexity is the study of higher order reactions, specifically termolecular reactions (i.e., third-order chemical reactions involving the mutual interaction between three reactants). Termolecular reactions play a crucial role in a multitude of fundamental physicochemical processes spanning more than 14 orders of magnitude in energy from plasma physics to ultracold atoms~\cite{SHUI197121,10.1093/mnrasl/slad047,Review_3B}. One of the most well-studied and relevant termolecular reactions is atom recombination, in which two atoms form a bound state while the third one carries away the excess kinetic energy~\cite{atom_recmbination1,atom_recombination2,atom_recombination3,Lindemann,Hinshelwood}. Even more interesting is the ion-atom-atom recombination reaction, which results in either the formation of a molecular ion or a molecule, depending on the collision energy~\cite{Krukow_2016,JPR_2018}. However, despite its relevance, no effort has been made to apply ML techniques to the prediction of these reactions. 

\begin{figure*}
    \centering
    \includegraphics[width=0.8\linewidth]{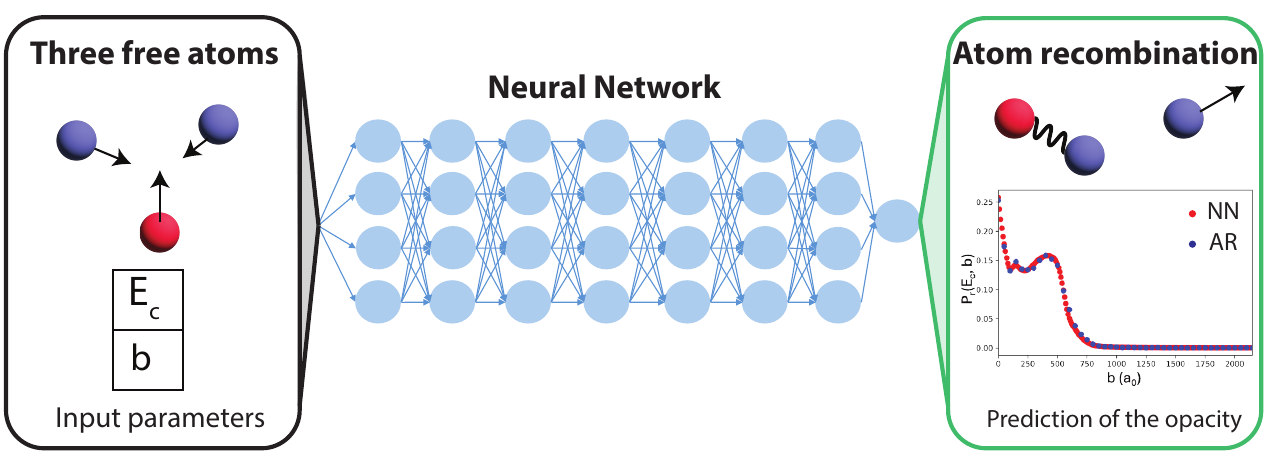}
    \caption{Displayed above is the overall scheme of this work. The collision energy $E_c$ and impact parameter $b$ are the model features, and the opacity function $P_r(E_c, \ b)$, which determines the probability of reaction, is the model target. The models developed in this work are meant to predict opacities for molecular formation resulting from the mutual collision of three atoms. The neural network shown here exactly depicts the architecture (layers and neurons) of the neural network models used in this work.}
    \label{NN_fig}
\end{figure*}

In this Letter, using a feed-forward neural network approach, we predict probabilities for ion-atom-atom recombination in an energy range between 10$^{-4}$ and 10$^4$~K with a $\lesssim 5\%$ relative error. The machine is capable of learning the reaction properties coming from long-range and short-range physics. Furthermore, it excellently predicts the maximum impact parameter for the reaction to occur--the essential element for the calculation of the reaction rate. The reaction rates are calculated from the ML-predicted reaction probabilities, yielding excellent agreement in both the low and high collision energy regimes. Moreover, the ML model can correctly extrapolate the reaction rates for collision energies well beyond the range explored in the training set, demonstrating that the machine can learn the reaction dynamics of atom recombination reactions.

To train any ML model to predict probabilities for atom recombination reactions, we must first obtain a detailed and representative training dataset. Here, the training dataset is obtained using the classical trajectory calculation method in hyperspherical coordinates, which has been shown to yield accurate results versus experimental data for a wide range of collision energies. Since the methodology has been previously introduced, we discuss it only briefly here~\cite{Review_3B}. The main idea is to map the three-body problem of three interacting particles of masses $m_i$ and positions \textbf{r}$_i$, $i=1,2,3$ into a single-particle problem in a 6D space. The three-body Hamiltonian is given by

\begin{equation}
 H = \frac{\textbf{p}_1^2}{2m_1} \ + \  \frac{\textbf{p}_2^2}{2m_2} \ \ + \ \frac{\textbf{p}_3^2}{2m_3} \ + \ V(\textbf{r}_1,\textbf{r}_2,\textbf{r}_3),
    \label{hamiltonian}
\end{equation}
and after neglecting the trivial center of mass motion, the Hamiltonian reads 
\begin{equation}
H = \frac{\textbf{P}_{6D}^2}{2\mu} + V(\boldsymbol{\rho}_{6D})
    \label{H_6D}
\end{equation}
where $\mu \ = \ \sqrt{\frac{m_1m_2m_3}{m_1 + m_2 + m_3}}$ is the reduced three-body mass, $\boldsymbol{\rho}_{6D}$ is the position in 6D configuration space and $\textbf{P}_{6D}$ is the conjugate momentum. In this space, it is straightforward to calculate the recombination cross section, resembling the standard two-body classical scattering cross section expression and given by 

\begin{equation}
\sigma_r(E_c) = \frac{8\pi^2}{3}\int_0^{b_{\text{max}}(E_c)}P_r(E_c, b)b^4db.
    \label{cross_sec}
\end{equation}
where $E_c$ is the collision energy, $b$ is the impact parameter, $b_{\text{max}}$ is the maximum impact parameter for which recombination occurs at a given $E_c$, and $P_r(E_c, b)$ is the so-called opacity function or reaction probability, representing the probability of a particular reaction as a function of $E_c$ and $b$. For a given $E_c$ and $b$, we run $n_{tot}$ trajectories and the products of the collision are tracked for each trajectory, yielding $n_r$ trajectories resulting in a reaction, hence 
\begin{equation}
P_r(E_c, b) = \frac{n_r}{n_{tot}} \pm \frac{\sqrt{n_r}}{n_{tot}}\sqrt{\frac{n_{tot}-n_r}{n_{tot}}},
    \label{opacity}
\end{equation}
where the uncertainty is a statistical result arising from the Monte-Carlo sampling technique taking into account that the reaction probability is a Boolean function. Finally, the energy-dependent atom recombination rate can be found by considering the volume (in 6D) of product formed per unit time

\begin{equation}
k_3(E_c) = \sqrt{\frac{2E_c}{\mu}}\sigma_r(E_c).
    \label{rate_eq}
\end{equation}

The reactants of choice are Sr$^+$ + Cs + Cs, whose reaction present two possible products: the molecular ion reaction channel Sr$^+$ + Cs + Cs $\rightarrow$ SrCs$^+$ + Cs and the molecule reaction channel Sr$^+$ + Cs + Cs $\rightarrow$ Sr$^+$ + Cs$_2$. Interestingly enough, all ion-atom-atom reactions present the same physics, and hence a single reaction is enough to check the viability of the machine learning approach to the cited reactions~\cite{Marjan2023}. We calculate the opacities for both reaction products in a range of energies between $ 10^{-4}$~K and $ 10^4$~K. We use a feed-forward neural network (FNN) to predict the opacity function. The FNN architecture used in this work (see Fig.~\ref{NN_fig}) is fully connected, which means the outputs of each neuron in a given layer are fed into every neuron in the following layer. The models were trained on the opacity functions generated from classical trajectory simulations. The features for the models were the collision energy $E_c$ and the impact parameter $b$~\footnote{specifically, the features were $\text{log}_{10}(E_c)$ with $E_c$ in K and $b$ in units of the Bohr radius a$_0$. The target of the models is the opacity function $P_r(E_c, b)$}. However, since some of the reaction probabilities can be very small, it proved better to learn $\sqrt{P_r(E_c, b)}$. Two models were trained, one for $\text{SrCs}^+$ formation and another for $\text{Cs}_2$ formation, and both had the same architecture, namely the one depicted in Fig.~\ref{NN_fig}. The model for $\text{SrCs}^+$ formation is referred to as NN-ion and the remaining model NN-mol. The models each have 137 tunable parameters (weights plus biases). The activation function used for all neurons except the lone neuron in the final layer was the softplus activation function. The final layer's neuron uses the sigmoid activation function, which ensures the outputs, meant to be (the square root of) probabilities, are in [0, 1].

The training dataset is generated from classical trajectories, using Lennard-Jones potentials assuming pairwise interactions, which has been shown to be an excellent approximation~\cite{Yu2024}. The parameters for the potentials (in atomic units) are: $C_6$ = 6.64 $\times 10^3$ and $C_{12}$ = 6.63 $\times 10^8$ for Cs-Cs, and $C_4$ = 200 and $C_8$ = 1.67 $\times 10^6$ for Cs-Sr$^+$ ~\cite{10.1063/1.3611399, 10.1063/1.448618, Schwerdtfeger18062019}. The classical trajectories were calculated using the software package Py3BR ~\cite{py3br}. For NN-ion, the energies used for training are $E_c \ \in [ 0.1  \text{mK}, \ 1  \text{mK} , \  10  \text{mK} , \ 0.1   \text{K}, \ 1 \text{K}, \ 10 \text{K}, \ 100 \text{K}, \ 1,000  \text{K}, $ \  $\ 5,000  \text{K}, \ 10,000  \text{K}]$, and we run 10$^4$ trajectories per impact parameter at all $E_c$. For NN-mol, the training energies are $E_c \ \in [10  \text{K}, \ 100   \text{K}, \ 1,000  \text{K}, \ 10,000  \text{K}]$, but 5$\times 10^4$ trajectories per impact parameter are required due to its typically small reaction probability. For all opacities in the training (or test) datasets, it was ensured that the maximum impact parameter $b_{max}$ was reached (or exceeded) by inspecting each opacity~\footnote{Furthermore, trajectories for additional impact parameters were run as needed to either reach $b_{max}$ or to more finely sample impact parameters, which were uniformly sampled for $b \ \in [0, \ b_{max}]$.}. For training, the Adam optimizer was used to fit the models' parameters with a learning rate of 0.007 ~\cite{kingma2014adam}. Glorot normal initialization was used to initialize the parameter values ~\cite{pmlr-v9-glorot10a}. The mean squared error loss function was used. NN-ion was trained using $\sim$20,000 epochs to a final loss of $\sim$$10^{-4}$, and NN-mol was trained using 10,000 epochs to a final loss of $\sim$1.3 $\times 10^{-5}$. The FNNs were designed and trained using the TensorFlow package and Keras API ~\cite{tensorflow2015-whitepaper}.

\begin{figure}
    \centering
    \includegraphics[width=0.7\linewidth]{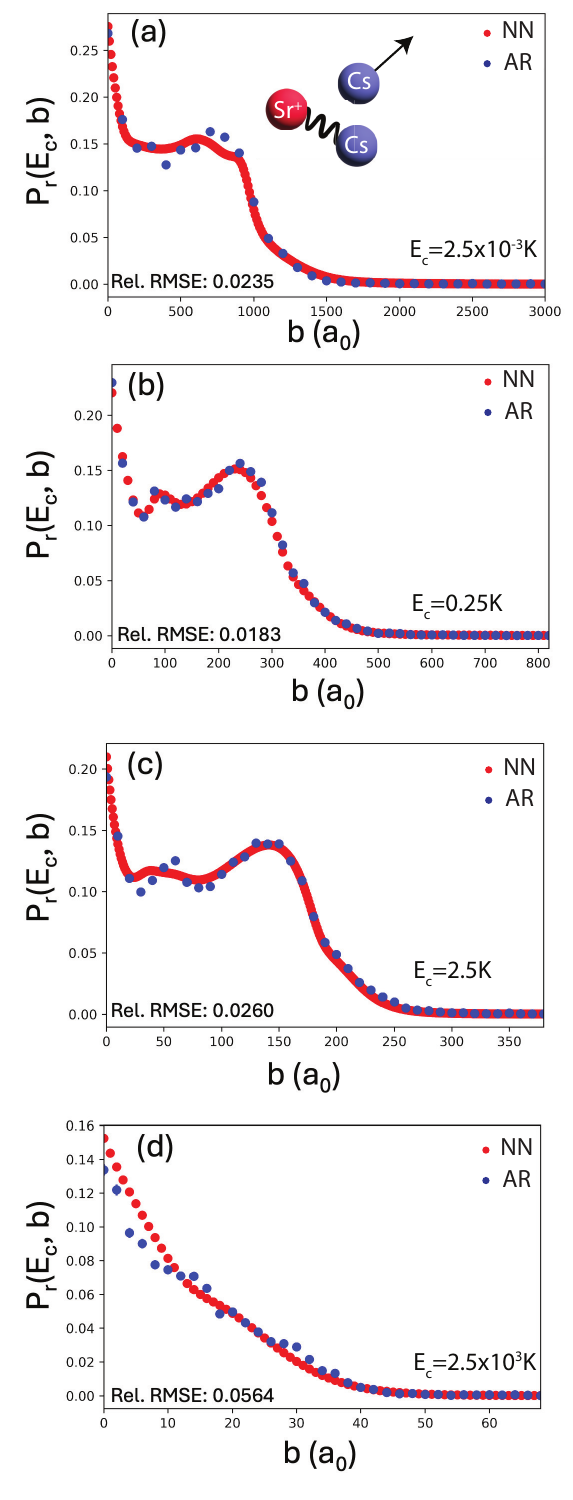}
    \caption{FNN-prediction of opacity functions for the formation of SrCs$^+$. Data points in red represent neural network predictions, and those in blue represent the results generated from classical trajectories (AR for atom recombination). Given also are the relative RMSE between predictions and simulation results.}
    \label{ion_opcities}
\end{figure}

The results showcasing the ability of the NN-ion model developed in this work are shown in Fig.~\ref{ion_opcities}. Each panel represents a different value of $E_c$ and shows the variation of the reaction probability with $b$. All of the values of $E_c$ used for the test results in Fig.~\ref{ion_opcities} are within the range of $E_c$ used for the model's training dataset. Also provided are the relative root-mean-squared error (RMSE) values comparing the NN predictions to the results from classical trajectory simulations (denoted "AR"). It is clear from Fig.~\ref{ion_opcities} that NN-ion demonstrates excellent predictive power over a wide range of energies. Indeed, NN-ion was trained on eight decades of $E_c$, and Fig.~\ref{ion_opcities} confirms that NN-ion makes accurate predictions throughout that energy range. All of the relative RMSE values in Fig. \ref{ion_opcities} are below 6\%, and NN-ion captures the opacity function throughout the entire range $b \in \ [0, \ b_{max}]$. Furthermore, it is clear, especially from panel (b), that NN-ion can predict the details of the variation of the opacity for small to intermediate values of $b$. Most importantly, the NN captures the change of the opacity function from the low-energy regime to the high energy regime around the dissociation energy of the molecular ion. In other words, the machine is able to differentiate short-range from long-range dominated atom recombination reactions, depending on the collision energy. Moreover, the NN is capable of locating $b_{max}$ for every collision energy.

\begin{figure}
    \centering
    \includegraphics[width=0.8\linewidth]{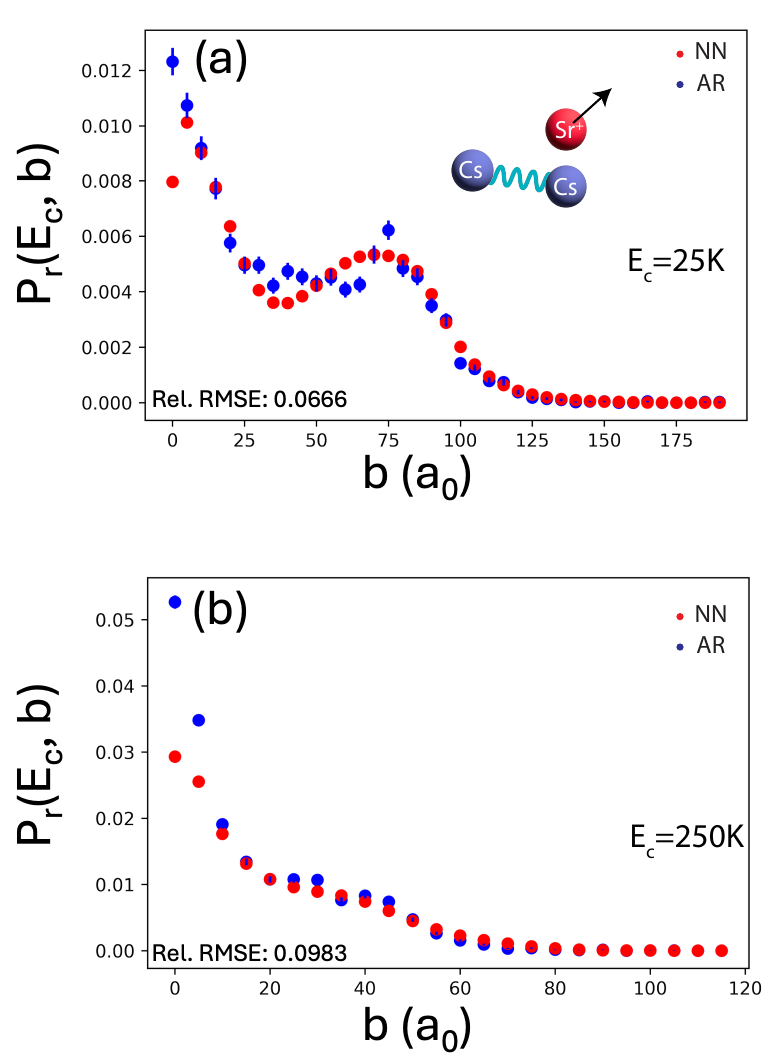}
    \caption{FNN-prediction of opacity functions for the formation of Cs$_2$. Data points in red represent neural network predictions, and those in blue represent the results generated from classical trajectories (AR for atom recombination). Given also are the relative RMSE between predictions and simulation results.}
    \label{mol_opacities}
\end{figure}

Representative results for NN-mol are given in Fig.~\ref{mol_opacities}, and as in Fig.~\ref{ion_opcities}, every panel stands for the opacity as a function of the impact parameter for a given collision energy. Specifically, panel (a) corresponds to an $E_c$ of 25K and panel (b) corresponds to an $E_c$ of 250K. This model shows a relative RMSE 10\%, coming mostly from the intermediate to small values of the impact parameters. However, as we show below, these errors will have little impact on the atom recombination rate since large impact parameters contribute the most. Thus, we may also note that for molecular formation, the model successfully learns the long-range behavior of the opacity function and anticipates $b_{max}$, two properties of the model that are essential for effective rate prediction. Furthermore, it is apparent from panel (a) that NN-mol roughly learns the details of the opacity function structure for small and intermediate values of $b$. It is clear that the NN approach to predicting opacity functions is universal between the reaction channels of an atom recombination reaction provided that separate models, having the same architecture, are trained on the different channels.

\begin{figure}[h]
    \centering
    \includegraphics[width=1.0\linewidth]{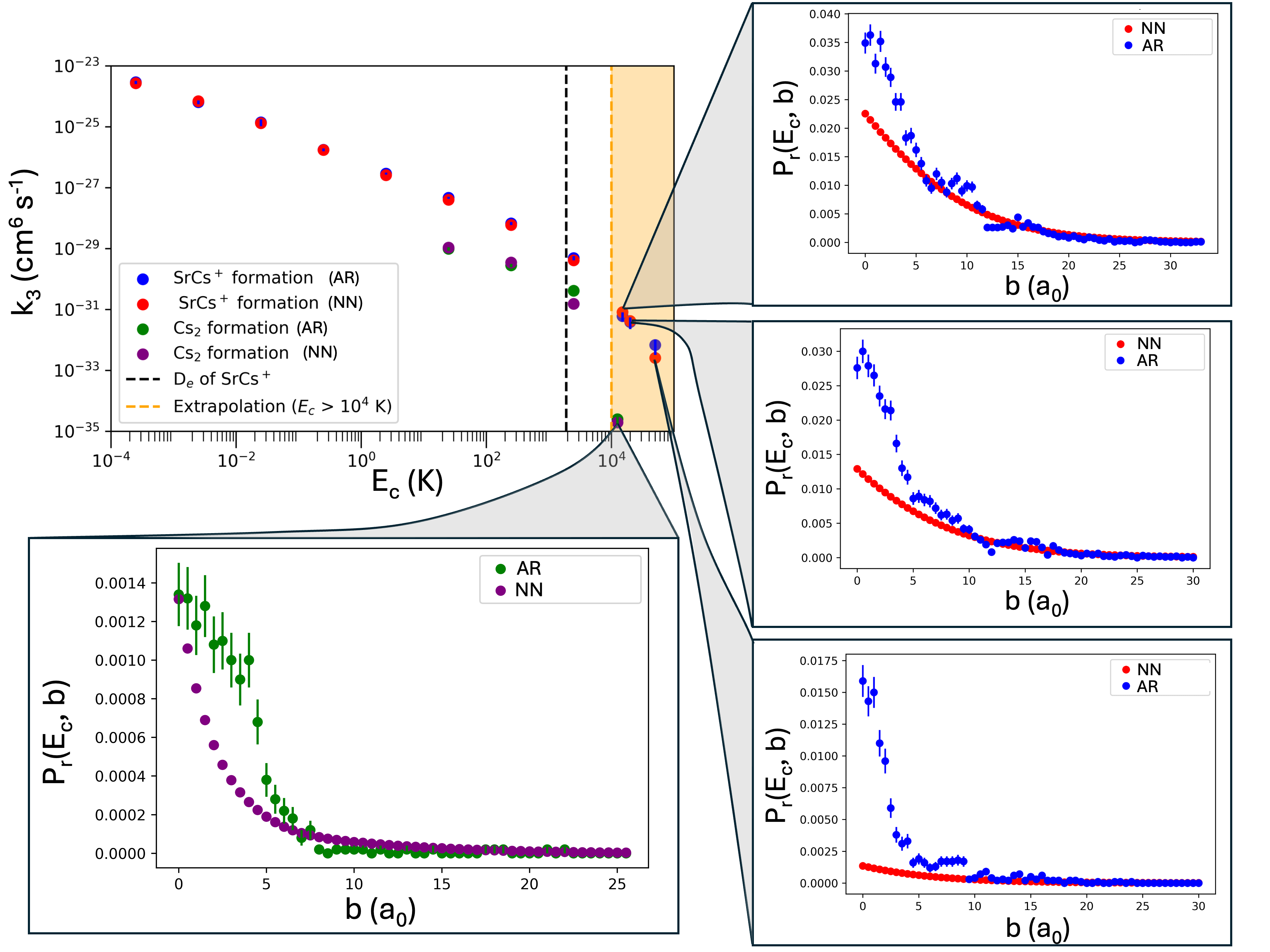}
    \caption{Energy-dependent atom recombination reaction rates k$_3$, for both SrCs$^+$ and Cs$_2$ formation for the collision energies used in the test datasets. In the k$_3$ plot, red points indicate SrCs$^+$ formation rates calculated from neural network (NN) predicted opacities, and those in blue indicate the rates calculated from classical trajectory simulations (AR for atom recombination). Also, Cs$_2$ formation is represented by purple points for those rates calculated from NN predicted opacities, and green points represent those rates calculated from classical trajectory simulations. The black, dashed line indicates where the threshold law takes effect, which corresponds to the dissociation energy of SrCs$^+$ ($D_e \approx 1,888$K ~\cite{py3br}). The orange-shaded region indicates the extrapolation regime where the models predicts opacities for collision energies greater than any included in the training dataset. }
    \label{rates}
\end{figure}

Here, we go beyond the interpolating regime by asking the machine for predictions in a totally unseen range of energies, very different from the ones the machine has been exposed too. Most machine learning methods have a difficult time with this task. Energy-dependent atom recombination reaction rates were calculated from both the classical trajectory simulations and the corresponding NN predictions for the test dataset, and the results are shown in Fig.~\ref{rates}, where the main panel shows the energy-dependent atom recombination rates as a function of the collision energy. At low collision energies, the reaction rate follows a power law until reaching the binding energy of the molecular ion (black-dashed line) where a new trend emerges. This change of trend is due to the role of the short-range region of the atom-ion interaction potential, establishing the onset of high energy collisions where the molecular formation channel starts to be comparable to molecular ion production. Therefore, for collision energies larger than the threshold, the predictions should be more involved due to the role of both short and long-range effects on the reaction dynamics. As displayed in Fig.~\ref{rates}, the NN-ion and NN-mol models are able to capture the regions of the opacity function having medium to large values of the impact parameter. As a result, the energy-dependent predicted reaction rates agree extremely well with the calculations.

Our work shows that machine learning techniques for the prediction of chemical reactions is readily applicable to high-order reaction rates such as atom recombination reactions. We show that machine learning models, specifically feed-forward neural networks, can efficiently learn atom recombination reactions in a collision energy range spanning from 10$^{-4}$ to 10$^{4}$~K, or equivalently, between 8.62$\times 10^{-9}$~eV to 8.62$\times 10^{-1}$~eV. We are able to predict the opacities for molecular ion recombination and molecule formation, showing the potential of machine learning techniques for termolecular reactions. Using ML predicted opacities, it is possible to calculate atom recombination cross sections and energy-dependent reaction rates. More spectacular is the fact that, even in the extrapolation regime where the collision energy is well beyond the greatest energy of the training dataset, both models perform extremely well, yielding very accurate reaction rates. The predicted reaction rates are accurate even when the model struggles to accurately capture the short impact parameter region. Alternatively, the machine is capable of differentiating short- from long-range effects as a function of the collision energy, showing that the machine learns the physics behind the opacity function. Our results open up a new avenue for machine learning techniques in the physical sciences: termolecular reactions. 

The authors acknowledge the support of the United States Air Force Office of Scientific Research [grant number FA9550-23-1-0202].


%

\end{document}